%% file: neurips_2025.tex
\documentclass{article}

\PassOptionsToPackage{numbers, compress}{natbib}

\usepackage[dblblindworkshop, final]{neurips_2025}

\workshoptitle{Fourth Workshop on Deep Learning for Code}

\usepackage[utf8]{inputenc} 
\usepackage[T1]{fontenc}    
\usepackage{hyperref}       
\usepackage{url}            
\usepackage{booktabs}       
\usepackage{amsfonts}       
\usepackage{nicefrac}       
\usepackage{microtype}      
\usepackage[table]{xcolor}
\usepackage{tcolorbox}

\usepackage{algorithm}
\usepackage{algorithmicx}
\usepackage[noend]{algpseudocode}
\usepackage{tabularx}

\usepackage{amsmath}
\usepackage{amssymb}
\usepackage{cleveref}
\usepackage{xspace}
\usepackage{multirow}
\usepackage{subcaption}

\usepackage{listings}

\title{\name: A Multi-Agent System for GPU Kernel Performance Optimization}

\author{
Anjiang Wei$^{1}$\thanks{Correspondence to: anjiang@cs.stanford.edu} \quad Tianran Sun$^{2}$ \quad Yogesh Seenichamy$^{1}$ \quad Hang Song$^{1}$ \\
\textbf{Anne Ouyang$^{1}$ \quad Azalia Mirhoseini$^{1}$ \quad Ke Wang$^{3}$ \quad Alex Aiken$^{1}$} \\
\\
$^{1}$Stanford University \quad
$^{2}$Shanghai Jiao Tong University \quad
$^{3}$Nanjing University \\
}

\begin{document}

\input{macros}

\maketitle

\begin{abstract}
GPU kernel optimization has long been a central challenge at the intersection of high-performance computing and machine learning. Efficient kernels are crucial for accelerating large language model (LLM) training and serving, yet attaining high performance typically requires extensive manual tuning. Compiler-based systems reduce some of this burden, but still demand substantial manual design and engineering effort. Recently, researchers have explored using LLMs for GPU kernel generation, though prior work has largely focused on translating high-level PyTorch modules into CUDA code. In this work, we introduce \name, the first LLM-based multi-agent system for GPU kernel optimization. Unlike previous approaches, \name{} starts from existing CUDA implementations extracted from SGLang, a widely deployed framework for serving LLMs, rather than treating PyTorch modules as the specification. Within \name{}, specialized LLM agents collaborate through iterative code generation, testing, profiling, and planning to produce kernels that are both correct and high-performance. On kernels from SGLang, \name{} achieves an average speedup of \avgspeedup using zero-shot prompting with OpenAI o4-mini. A detailed case study further demonstrates that LLMs can autonomously apply loop transformations, optimize memory access patterns, exploit CUDA intrinsics, and leverage fast math operations to yield substantial performance gains. Our work highlights multi-agent LLM systems as a promising new paradigm for GPU kernel optimization. Our code is publicly available at \url{https://github.com/Anjiang-Wei/Astra}.
\end{abstract}

\input{1_intro}
\input{2_related}
\input{3_method}
\input{4_setup}
\input{5_result}
\input{6_discussion}
\input{7_conclusion}







\bibliographystyle{IEEEtran}
\bibliography{custom}


\end{document}

%% file: macros.tex
\newcommand{\anjiang}[1]{\textcolor{blue}{\textbf{Anjiang:} #1}}

\newcommand{\ke}[1]{\textcolor{red}{\textbf{Ke:} #1}}

\newcommand{\joy}[1]{\textcolor{blue}{\textbf{Joy:} #1}}

\newcommand{\hang}[1]{\textcolor{blue}{\textbf{Hang:} #1}}

\newcommand{\alex}[1]{\textcolor{blue}{\textbf{Alex:} #1}}
\newcommand{\anne}[1]{\textcolor{pink}{\textbf{Anne:} #1}}

\newcommand{\CodeIn}[1]{\texttt{#1}}

\newcommand{\maxspeedup}{1.46×\xspace}
\newcommand{\avgspeedup}{1.32×\xspace}
\newcommand{\singlespeedup}{1.08×\xspace}

\newcommand{\name}{Astra\xspace}

\newcommand{\merge}{\texttt{merge\_attn\_states\_lse}\xspace}
\newcommand{\rmsnorm}{\texttt{fused\_add\_rmsnorm}\xspace}
\newcommand{\silu}{\texttt{silu\_and\_mul}\xspace}

\definecolor{codekw}{RGB}{0,0,150}   
\definecolor{codecm}{gray}{0.45}     

\lstdefinelanguage{CUDA}{
  language=C++,
  morekeywords={
    __global__,__device__,__host__,__shared__,__constant__,__managed__,
    __inline__,__forceinline__,__restrict__,
    __syncthreads, __syncwarp, __threadfence, __threadfence_block, __threadfence_system,
    __ldg, __shfl_sync, __activemask, __ballot_sync,
    warpSize, threadIdx, blockIdx, blockDim, gridDim,
    dim3, uint2, int2, float2, float4, __half, __half2,
    make_uint2, make_int2, make_float2, make_float4
  },
  morekeywords={
    cudaMalloc, cudaFree, cudaMemcpy, cudaMemset, cudaDeviceSynchronize,
    cudaGetLastError, cudaMemcpyAsync, cudaStreamCreate, cudaStreamDestroy,
    cudaOccupancyMaxActiveBlocksPerMultiprocessor
  },
  alsoletter=<>,
  morekeywords={<,>},
  sensitive=true,
  comment=[l]{//},
  commentstyle=\color{codecm}\ttfamily,
  morecomment=[s]{/*}{*/},
  stringstyle=\ttfamily, 
  basicstyle=\ttfamily\footnotesize,
  keywordstyle=\bfseries\color{codekw},
  identifierstyle=\color{black},
  numbers=left,
  numberstyle=\tiny\color{codecm},
  stepnumber=1,
  numbersep=6pt,
  showstringspaces=false,
  keepspaces=true,
  columns=fullflexible,
  tabsize=2,
  breaklines=true,
  frame=single,
  rulecolor=\color{black}, 
  aboveskip=4pt,
  belowskip=4pt,
  captionpos=b
}

\lstdefinestyle{cuda-compact}{
  language=CUDA,
  basicstyle=\ttfamily\footnotesize,
  numbers=left,
  frame=single,
  rulecolor=\color{black}, 
  breaklines=true,
  columns=fullflexible,
  keepspaces=true,
  showstringspaces=false,
  aboveskip=4pt,
  belowskip=4pt,
  numbersep=6pt
}

\lstdefinestyle{cuda-tight}{
  language=CUDA,
  basicstyle=\ttfamily\scriptsize,
  numbers=left,
  frame=single, rulecolor=\color{black},
  showstringspaces=false,
  keepspaces=true, columns=fullflexible,
  breaklines=false,     
  xleftmargin=0pt, framexleftmargin=0pt,
  aboveskip=2pt, belowskip=2pt,
  basewidth=0.52em      
}

%% file: 1_intro.tex
\section{Introduction}
\label{sec:intro}

 Recent advances in large language models (LLMs) have led to state-of-the-art performance on a wide range of tasks, including reasoning and code generation~\cite{chen2021evaluating,austin2021program,wei2025vericoder,hendrycks2021measuring,ahn2024large,wei2025satbench}. Building on these capabilities, autonomous agents powered by LLMs have begun to automate parts of the software development pipeline~\cite{jimenez2023swe,jain2024r2e,yang2024swe}. In this work, we investigate the application of LLM-powered agents to GPU kernel optimization, a long-standing challenge at the intersection of high-performance computing and machine learning that requires generating code that is both correct and highly optimized.

GPU kernel optimization is essential for improving the efficiency of LLM serving and training, which is critical for the successful deployment of LLMs. However, even with decades of advances in GPU programming, kernel development remains a fundamentally difficult real-world engineering problem. Rapid hardware evolution often requires extensive manual tuning and reimplementation. For example, FlashAttention-2~\cite{dao2023flashattention} suffered a 47\% performance drop when first ported to NVIDIA’s H100 GPUs, and it was only after more than two years that FlashAttention-3~\cite{shah2024flashattention} introduced substantial new optimizations to recover performance. In addition, emerging model architectures~\cite{gu2023mamba,ho2020denoising,li2022diffusion,liu2024deepseek} and dynamic workloads with variable input lengths~\cite{sun2024llumnix} further complicate kernel optimization. As a result, many available kernel implementations operate well below hardware peak. Closing this gap is vital for advancing performance, reducing costs, and improving energy efficiency.

From a systems perspective, there are two dominant paradigms for GPU kernel optimization. The first is fully manual tuning, exemplified by libraries such as NVIDIA cuDNN~\cite{chetlur2014cudnn}. This approach demands extensive manual effort, involves time-consuming engineering cycles, and can still leave optimization opportunities untapped. The second paradigm is compiler-based optimization, represented by systems such as TVM~\cite{chen2018tvm}, Triton~\cite{tillet2019triton}, Mirage~\cite{wu2024multi}, ThunderKittens~\cite{spector2024thunderkittens}, and others~\cite{zheng2020ansor,zheng2022amos}. For instance, Triton introduces a tile-level intermediate representation combined with autotuning to deliver performance close to hand-optimized kernels, significantly reducing the engineering burden for end users. Nevertheless, these compiler-based systems themselves require substantial engineering effort to develop and must be continuously adapted as hardware evolves.

Given the significant potential of LLMs, researchers have actively explored their use for GPU kernel optimization. KernelBench~\cite{ouyang2025kernelbench} is a pioneering work that first formulates the task for LLMs and introduces a corresponding benchmark. Other studies have explored single-agent approaches~\cite{lange2025aicudaengineer,chen2025cudallmllmswriteefficient,andrews2025gpukernelscientistllmdriven} as well as training-based methods for improving LLMs~\cite{baronio2025kevinmultiturnrlgenerating,li2025cudal1improvingcudaoptimization}.

\begin{figure}
    \centering
    \includegraphics[width=\linewidth]{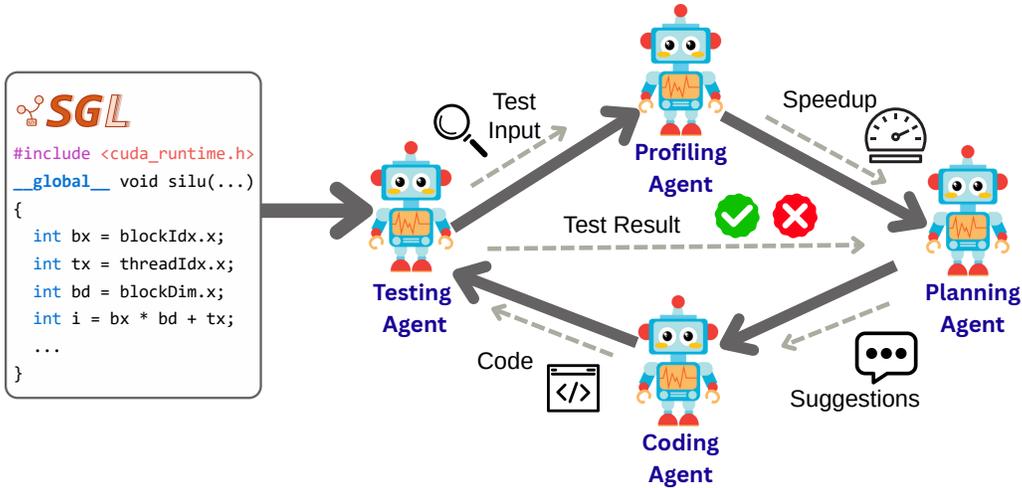}
    \caption{Overview of \name. Given an existing GPU kernel extracted from SGLang, \name{} employs a multi-agent approach for kernel optimization, where specialized LLM agents collaborate through iterative code generation, testing, profiling, and planning to produce correct and high-performance kernels.}
    \label{fig:intro}
\end{figure}

In this work, we introduce \name, the first LLM-based multi-agent system for GPU kernel optimization. Our key observation is that kernel optimization is inherently a multi-stage process that includes code generation, testing, profiling, and planning, and a single LLM agent is unlikely to excel at all of these tasks. As shown in \Cref{fig:intro}, \name{} addresses this challenge by decomposing the task into specialized agents that collaborate iteratively. This coordinated workflow leverages complementary agent capabilities, enabling systematic exploration of the optimization space and consistently producing kernels that are both correct and high-performance.

Our setting contrasts with KernelBench~\cite{kernelbench} in two important respects. Unlike KernelBench, which frames the task as generating CUDA kernels from high-level PyTorch models written in Python, we focus on optimizing existing CUDA implementations. This reflects the reality of production environments, where kernels are already available and the real challenge is squeezing out additional performance rather than generating CUDA code from scratch. KernelBench has already demonstrated that translation from Python to CUDA is non-trivial for LLMs; our work focuses squarely on performance optimization and avoids the additional burden of translation, which can introduce errors and degrade performance. In addition, our kernels are taken directly from SGLang~\cite{zheng2025sglang} and can be seamlessly reintegrated into the system. As SGLang is a production-grade LLM serving framework deployed at scale and responsible for generating trillions of tokens per day across major enterprises and institutions, even modest kernel-level improvements can yield substantial real-world impact.

We evaluate \name{} on three kernels extracted from SGLang~\cite{zheng2025sglang} and observe an average speedup of \avgspeedup{} using zero-shot prompting with OpenAI o4-mini. Importantly, these results are achieved without any additional training, including supervised fine-tuning or reinforcement learning, which highlights the effectiveness of our approach in a pure prompting setting and suggests further potential when combined with training-based methods. To demonstrate the necessity of dedicated agent roles, we compare against a single-agent baseline, which attains only \singlespeedup{} speedup on average. Finally, we conduct an in-depth case analysis to investigate the source of performance gains. Our findings show that LLMs can autonomously apply loop transformations, restructure memory access patterns, make extensive use of CUDA intrinsics, and exploit fast math operations, all of which contribute to the observed speedups.

In summary, our contributions are:
\begin{itemize}
\item We design and implement \name{}, a multi-agent system for GPU kernel optimization, in which specialized LLM agents collaborate through iterative code generation, testing, profiling, and planning to produce correct and high-performance kernels.
\item We demonstrate an average speedup of \avgspeedup on kernels from SGLang, a production-grade LLM serving framework, and our optimized kernels can be seamlessly reintegrated to deliver substantial real-world impact.
\item We conduct a detailed manual analysis of the kernels generated by \name{} and identify the optimization strategies, including loop transformations, memory access improvements, extensive use of CUDA intrinsics, and faster math operations, that account for the observed speedups.
\end{itemize}

%% file: 2_related.tex
\section{Related Work}
\label{sec:related}

\paragraph{Multi-Agent Systems}

Multi-agent systems (MAS) consist of multiple interacting agents that collaborate to solve complex, shared problems that exceed the capabilities of a single agent. This paradigm is particularly well-suited to programming, where intricate workflows can be naturally decomposed into sub-tasks such as planning, implementation, testing, and profiling. Recent work has explored multi-agent frameworks including AutoGen~\cite{wu2024autogen}, Trace~\cite{cheng2024trace}, and MetaGPT~\cite{hong2023metagpt,hong2024metagpt}, which have demonstrated strong performance on benchmarks in mathematics and code generation~\cite{li2023camel,huang2024agentcodermultiagentbasedcodegeneration,qian2024chatdevcommunicativeagentssoftware,wei2025improving,yang2024swe}. However, there has been little exploration of applying MAS to GPU kernel optimization, a domain where highly specialized performance considerations introduce unique challenges.

\paragraph{Compiler and Learning-Based Approaches to GPU Kernel Optimization}

GPU kernel optimization has long been driven by compiler frameworks and domain-specific languages (DSLs). Systems such as Halide~\cite{ragan2013halide}, TVM~\cite{chen2018tvm}, MLIR~\cite{lattner2020mlircompilerinfrastructureend}, TensorFlow XLA~\cite{abadi2016tensorflow}, and NVIDIA CUTLASS~\cite{CUTLASS}, along with others~\cite{rotem2018glow,wei2017dlvm,jia2019beyond,zheng2021neoflow,wei2025task}, provide high-level abstractions for expressing tensor computations and support compiler-driven optimizations. To further improve performance, autotuning frameworks such as AutoTVM~\cite{chen2018tvm}, Ansor~\cite{zheng2020ansor}, and AMOS~\cite{zheng2022amos} leverage search and machine learning to explore large optimization spaces. More recent systems, including Triton~\cite{tillet2019triton}, Mirage~\cite{wu2024multi}, and ThunderKittens~\cite{spector2024thunderkittens}, expand on these ideas. For example, Triton introduces a tile-level intermediate representation and autotuning, achieving performance close to hand-optimized kernels. Nevertheless, compiler-based approaches often fall short of expert-level performance without extensive tuning, and generalization across hardware platforms remains difficult~\cite{spector2024thunderkittens}. Despite their progress, these systems are still constrained by rigid compilation pipelines and require significant engineering effort to build.

\paragraph{LLM-Driven Approaches to High-Performance Code Generation}
Early efforts in LLM-based code generation, such as AlphaCode~\cite{li2022competition}, primarily targeted general-purpose programming tasks and demonstrated promising results. More recently, research has increasingly focused on domain-specific high-performance code generation, spanning tasks such as vectorization~\cite{taneja2025llm,zheng2025vectransenhancingcompilerautovectorization}, assembly-level optimization~\cite{wei2025improvingassemblycodeperformance}, parallel programming with domain-specific languages (DSLs)~\cite{wei2024improving,wei2025improving,wei2025mapple}, and tensor program optimization~\cite{zhai2024enabling}. A particularly active direction is the automatic generation of performant GPU kernels~\cite{ouyang2025kernelbench,li2025tritonbenchbenchmarkinglargelanguage}.
Because code optimization provides verifiable rewards, iterative refinement has emerged as a natural paradigm: models generate candidate kernels and progressively improve them through feedback loops involving compilation checks, correctness validation, runtime profiling, or self-reflection~\cite{ouyang2025kernelbench,chen2025automating,wei2025improving}. Unlike general code generation tasks, a central challenge in code optimization is to ensure that LLMs generate code that is both functionally correct and highly optimized, where correctness means equivalence to the original program for all inputs~\cite{wei2025equibench,wei2025equibench2}. To tackle this challenge, researchers have explored both prompt-based approaches~\cite{agrawal2025gepareflectivepromptevolution,novikov2025alphaevolvecodingagentscientific,chen2025cudallmllmswriteefficient,andrews2025gpukernelscientistllmdriven,lange2025aicudaengineer,ouyang2025surprisinglyfast} and training-based methods, including multi-turn reinforcement learning~\cite{baronio2025kevinmultiturnrlgenerating} and contrastive reinforcement learning~\cite{li2025cudal1improvingcudaoptimization}. Our work addresses this challenge by adopting a multi-agent system approach.

%% file: 3_method.tex
\section{Method}
\label{sec:method}
\subsection{Task Definition}
\label{subsec:task}

The goal of CUDA optimization is to produce an optimized kernel $S'$ that runs faster than the baseline kernel $S$ while preserving its functional correctness. Below, we formally define the correctness and performance criteria, and then outline how our setup differs from prior work.

\paragraph{Correctness.}
Let $\mathcal{X}$ be the input domain and $\mathcal{Y}$ the output space.
The baseline and optimized kernels are functions $S, S' : \mathcal{X} \to \mathcal{Y}$.
Ideally, we require
\[
\forall x \in \mathcal{X}:\; S'(x) = S(x),
\]
or, allowing floating-point deviations,
\[
\forall x \in \mathcal{X}:\; d\bigl(S'(x), S(x)\bigr) \le \varepsilon,
\]
for a discrepancy metric $d$ and tolerance $\varepsilon \ge 0$.
Since exact equivalence is undecidable in practice, we evaluate correctness on a finite test suite
\[
T = \{(x_i, y_i)\}_{i=1}^m,\qquad y_i := S(x_i),
\]
where the $x_i$ are chosen to represent diverse tensor shapes and values.
We deem $S'$ correct if
\[
\max_{1 \le i \le m} d\bigl(S'(x_i), y_i\bigr) \le \varepsilon.
\]

\paragraph{Performance.}
Let $\tau(S,x)$ denote the runtime of kernel $S$ on input $x \in \mathcal{X}$. 
For each input $x$, the speedup is
\[
\sigma(x) = \frac{\tau(S,x)}{\tau(S',x)}.
\]

To summarize results over the test suite $T$, we report the geometric mean $\sigma_T$, which is the standard choice for averaging speedups because it correctly aggregates ratios, is symmetric between speedups and slowdowns, and reduces the influence of outliers:
\[
\sigma_T = \Biggl(\prod_{i=1}^{m} \frac{\tau(S,x_i)}{\tau(S',x_i)}\Biggr)^{1/m}.
\]
The optimization objective is to maximize this geometric-mean speedup while preserving correctness.

\subsection{Multi-Agent System}
\label{subsec:system}

\paragraph{Agent Roles.}
As shown in \Cref{fig:intro}, \name{} is organized around four specialized agents, each responsible for a distinct stage of the CUDA optimization pipeline.
The \emph{testing agent} creates a suite of test cases from the baseline kernel and checks the correctness of candidate kernels.
The \emph{profiling agent} measures execution time on the test suite, providing performance feedback.
The \emph{planning agent} combines correctness and performance signals to propose targeted modifications.
The \emph{coding agent} applies these suggestions to generate new kernel implementations.
Together, these agents form a feedback loop that supports iterative refinement while preserving correctness.

\paragraph{Algorithm.}
\Cref{alg:alg} outlines the multi-agent optimization procedure. The process begins with the construction of an initial test suite and profiling of the baseline kernel. The system then proceeds through $R$ iterative rounds: in each round, the planning agent proposes modifications, the coding agent generates a new candidate kernel, and the testing and profiling agents re-evaluate correctness and performance. All results are recorded in a log of tuples $(\textit{round}, \textit{code}, \textit{correctness}, \textit{performance})$, where \textit{correctness} is a binary indicator of whether the candidate passes the tests. This log enables systematic tracking of the optimization trajectory.

\algrenewcommand\algorithmicrequire{\textbf{Input:}}
\algrenewcommand\algorithmicensure{\textbf{Output:}}
\algnewcommand\algorithmicdefine{\textbf{Define:}}
\newcommand\Define{\item[\algorithmicdefine]}
\algrenewcommand\algorithmicforall{\textbf{for each}}
\renewcommand{\textproc}{\texttt}

\newcommand{\var}[1]{\ensuremath{\mathsf{#1}}}

\begin{algorithm}[!tb]
\caption{Multi-Agent CUDA Optimization}
\label{alg:multiagent-cuda}
\begin{algorithmic}[1]
  \Require Baseline CUDA code $S_0$, number of rounds $R$
  
  \Define \texttt{TestingAgent} \Comment{Generate or run tests.}
  \Statex \hspace{\algorithmicindent}\hspace{0.3em}\texttt{ProfilingAgent} \Comment{Measure performance.}
  \Statex \hspace{\algorithmicindent}\hspace{0.3em}\texttt{PlanningAgent} \Comment{Propose suggestions given correctness and performance signals.}
  \Statex \hspace{\algorithmicindent}\hspace{0.3em}\texttt{CodingAgent} \Comment{Apply suggestions to previous code $S_{\text{prev}}$}
  \Statex \hspace{\algorithmicindent}\hspace{0.3em}\texttt{Log} \Comment{List of $(\textit{round}, \textit{code}, \textit{correctness}, \textit{performance})$ for all iterations.}
  \Statex \hspace{\algorithmicindent}\hspace{0.3em}\texttt{Test suite $T$} \Comment{Tests generated by \texttt{TestingAgent}.}
  
  \Ensure \texttt{Log}
  
\State $T \gets \Call{TestingAgent.GenerateTests}{S_0}$ \Comment{Initialization}
\State $\var{perf}_0 \gets \Call{ProfilingAgent.Profile}{S_0, T}$
\State $\var{Log} \gets [\,]$
\State \Call{Append}{\var{Log}, $(0, S_0, \var{True}, \var{perf}_0)$}
\State $S_{\text{prev}} \gets S_0$
\State $\var{pass}_{\text{prev}} \gets \var{True}$
\State $\var{perf}_{\text{prev}} \gets \var{perf}_0$

\For{$r \gets 1$ to $R$} \Comment{Iterative optimization starts}
  \State $\var{suggestions} \gets \Call{PlanningAgent.Suggest}{S_{\text{prev}}, \var{pass}_{\text{prev}}, \var{perf}_{\text{prev}}}$
  \State $S_{\text{new}} \gets \Call{CodingAgent.Apply}{S_{\text{prev}}, \var{suggestions}}$
  \State $\var{pass}_{\text{new}} \gets \Call{TestingAgent.Validate}{S_{\text{new}}, T}$
  \State $\var{perf}_{\text{new}} \gets \Call{ProfilingAgent.Profile}{S_{\text{new}}, T}$
  \State \Call{Append}{\var{Log},\; $(r, S_{\text{new}}, \var{pass}_{\text{new}}, \var{perf}_{\text{new}})$}
  \State $S_{\text{prev}} \gets S_{\text{new}}$
  \State $\var{pass}_{\text{prev}} \gets \var{pass}_{\text{new}}$
  \State $\var{perf}_{\text{prev}} \gets \var{perf}_{\text{new}}$
\EndFor
  
  \State \Return $\textit{Log}$
\end{algorithmic}
\label{alg:alg}
\end{algorithm}

\paragraph{Pre-Processing and Post-Processing.}
Allowing \name{} to directly optimize the raw CUDA kernels in the SGLang framework~\cite{zheng2025sglang} is difficult because these kernels have many internal dependencies. To address this, we perform a manual pre-processing step: extracting and simplifying the kernels into stand-alone versions that serve as the baseline inputs for \name. After optimization, we apply a post-processing step that integrates the generated kernels back into SGLang and validates them against the original framework implementation (rather than only the extracted version). We report speedups relative to the original SGLang kernels, ensuring that the optimized kernels can be seamlessly integrated into the framework as drop-in replacements and that performance is measured within the full framework.

%% file: 4_setup.tex
\section{Experimental Setup}
\label{sec:setup}



\paragraph{Metrics.} Our evaluation focuses on both correctness and performance. Correctness is determined using test cases that we construct with diverse tensor input shapes. We compare the outputs of generated kernels against the execution results of the original SGlang implementation, which serves as the ground truth. For performance, we measure the execution time of both the original kernel and the optimized version on the same tensor shapes, and report speedup as the metric. While the multi-agent framework internally produces its own test cases through the testing agent, the final evaluation relies on manually designed test cases to ensure high confidence in functional validation.

\paragraph{Kernels.} We evaluate three kernels from the LLM serving framework SGLang~\cite{zheng2025sglang}: \silu, \rmsnorm, and \merge. Their computations are summarized in \Cref{tab:kernels}.

\paragraph{Performance Measurement.}
We evaluate performance across a range of input shapes and report average results. For each input shape, we run 100 repetitions after 20 warm-up runs. The input shapes are selected based on the actual dimensions used in modern LLMs, including the LLaMA-7B, 13B, and 70B models. A detailed analysis of how input shapes affect performance is provided in \Cref{subsec:shape}.

\paragraph{Implementation.}
We implement our multi-agent system with the OpenAI Agents SDK framework~\cite{opensdk}, which offers standardized abstractions for defining agents and integrating function tools. The agents are powered by OpenAI’s \texttt{o4-mini} model, and all experiments are conducted on a machine equipped with NVIDIA H100 GPUs. We set the number of rounds to optimize $R$ to be 5.

%% file: 5_result.tex
\section{Results}
\label{sec:result}

\subsection{Main Results}
\label{subsec:perf}

\begin{table*}[!tb]
\centering
\small
\begin{tabularx}{\textwidth}{p{0.15\textwidth} p{0.30\textwidth} p{0.33\textwidth}}
\toprule
\textbf{Index} & \textbf{Kernel Name} & \textbf{Computation} \\
\midrule
Kernel 1 & \merge &
$\begin{array}{l}
\mathbf V_{\text{out}} = \dfrac{e^{S_a}\,\mathbf V_a + e^{S_b}\,\mathbf V_b}{e^{S_a} + e^{S_b}}, \\[0.4em]
S_{\text{out}} = \log\!\left(e^{S_a} + e^{S_b}\right)
\end{array}$ \\
\addlinespace
\hline
\addlinespace
Kernel 2 & \rmsnorm &
$\begin{array}{l}
\mathbf{y} = \dfrac{\mathbf{x}+\mathbf{r}}{\sqrt{\tfrac{1}{D}\|\mathbf{x}+\mathbf{r}\|_2^2+\varepsilon}} \odot \mathbf{w}
\end{array}$ \\
\addlinespace
\hline
\addlinespace
Kernel 3 & \silu &
$\begin{array}{l}
\mathbf{out} = \operatorname{SiLU}(\mathbf{x}) \odot \mathbf{g}, \\[0.4em]
\operatorname{SiLU}(z) = \tfrac{z}{1 + e^{-z}}
\end{array}$ \\
\bottomrule
\end{tabularx}
\caption{Kernel names and computations.}
\label{tab:kernels}
\end{table*}

\newcommand{\cmark}{\textcolor{green!60!black}{\checkmark}}

\begin{table}[!tb]
\centering
\begin{tabular}{lccc|cc>{\columncolor{gray!20}}c|c}
\toprule
Kernel & LoC-Base & LoC-Opt. & $\Delta$LoC & Time-Base & Time-Opt. & {\cellcolor{white}\textbf{Speedup}} & \textbf{Correct} \\
\midrule
Kernel 1 & 124 & 232 & +87\% & 31.4 &  24.9  & 1.26× & \cmark \\
Kernel 2 & 108 & 163  & +50\% & 41.3  &   33.1  & 1.25× & \cmark \\
Kernel 3 & 99 & 157 & +59\% & 20.1  & 13.8 & \maxspeedup & \cmark \\
\midrule
Average  & 110 & 184 & +64\% & 30.9  & 23.9 & \avgspeedup
& \cmark \\
\bottomrule
\end{tabular}
\vspace{1em}
\caption{Baseline vs. optimized kernels: Lines of Code (LoC) and execution time ($\mu s$). All kernels optimized by our multi-agent system are correct.}
\label{tab:speedup}
\end{table}

\paragraph{Correctness.}
As shown in the last column of \Cref{tab:speedup}, all three optimized kernels are validated against the original SGLang implementations and confirmed to be correct. As described in \Cref{sec:setup}, we do not rely on test cases generated by the testing agent for functional validation. Instead, we manually construct test cases for the kernels produced by \name and check their outputs against the original SGLang kernels.

\paragraph{Performance.}
\Cref{tab:speedup} summarizes the performance gains achieved by \name across the three kernels. The results show that \name can consistently improve performance while preserving correctness. For \merge (Kernel 1), the optimized version has 87\% more lines of code and delivers a 1.26$\times$ speedup. For \rmsnorm (Kernel 2), the optimized kernel contains 50\% more lines and achieves a 1.25$\times$ speedup. For \silu (Kernel 3), the optimized kernel has 59\% more lines and yields a 1.46$\times$ speedup. Overall, with only five optimization rounds, \name achieves an average speedup of \avgspeedup and up to \maxspeedup, measured over a set of representative tensor shapes. We present detailed case studies in \Cref{subsec:case} and analyze how tensor shapes influence performance in \Cref{subsec:shape}.

\subsection{Comparison with Single-Agent Approach}

\paragraph{Setup of Single-Agent Method.}
In the single-agent setting, we continue to use the OpenAI Agents SDK framework but instantiate only one agent. This agent handles all tasks, including testing, profiling, planning, and code generation, and has access to the same set of tools as in the multi-agent setting. For fairness, we run the same number of optimization rounds, set to five, and the only difference lies in the number of agents involved.

\paragraph{Performance.}
As shown in \Cref{tab:sa-ma-comparison}, the multi-agent approach achieves higher performance speedup than the single-agent approach (\avgspeedup vs. 1.08×), while both approaches consistently generate correct kernels. We observe that the advantages of the multi-agent setup become more pronounced as kernel complexity increases. For kernel 3, which is relatively simple, the performance of both approaches is comparable. In contrast, kernel 1 is the most complex and exposes the limitations of the single-agent setup, where certain tasks may not be carried out effectively enough to yield good overall results. In particular, the slowdown of Kernel 1 under the single-agent setting was due to unrepresentative test inputs generated during test construction, which biased the profiling results. This issue does not occur in the multi-agent approach, where one agent is dedicated to generating representative test inputs and another to conducting profiling. Overall, these findings demonstrate that \name provides greater advantages over the single-agent setup when handling more complex kernels.

\subsection{Case Studies}
\label{subsec:case}

\begin{table}[!tb]
\centering
\begin{tabular}{lccccc}
\toprule
Kernel & Time-Base ($\mu s$) & Correct - SA & Speedup - SA & Correct - MA & Speedup - MA \\
\midrule
Kernel 1 & 31.4 & \cmark & 0.73× & \cmark & 1.26× \\
Kernel 2 & 41.3 & \cmark & 1.18× & \cmark & 1.25×  \\
Kernel 3 & 20.1 & \cmark  & 1.48× & \cmark & 1.46× \\
\midrule
Average & 30.9 & \cmark & 1.08× & \cmark & \avgspeedup \\
\bottomrule
\end{tabular}
\vspace{1em}
\caption{Single-Agent (SA) vs. Multi-Agent (MA) comparison: baseline runtime (Time-Base), correctness, and speedup ($\times$).}
\label{tab:sa-ma-comparison}
\end{table}

We compare the source code of the baseline and \name{}-optimized kernels and conduct detailed performance profiling with NVIDIA Nsight Compute. Overall, the speedups stem from eliminating redundant computation, improving memory-access efficiency, and exploiting advanced CUDA features. Concretely, the optimized kernels apply loop transformations to enhance parallelism, adopt more aggressive memory-access strategies to maximize bandwidth, make extensive use of CUDA intrinsics for hardware-level efficiency, and leverage fast math operations. The following examples illustrate these optimization strategies.

\paragraph{Kernel 1: \merge}
A key optimization shown in \Cref{fig:merge-hoist} is hoisting loop-invariant computations out of the inner element loop. In the baseline, the mixing weights and their normalization are recomputed for every element of the output vector, incurring repeated exponentials and a division within the hot loop. The optimized version computes these quantities once per output vector, leaving the inner loop with only memory loads, multiply–add, and a store. By removing expensive operations from the loop body, the optimized kernel lowers the instruction count and increases throughput without affecting correctness.

\begin{figure}[t]
  \centering
  \begin{subfigure}[t]{0.47\columnwidth}
\begin{lstlisting}[style=cuda-compact,basicstyle=\ttfamily\scriptsize]
// two scalar scores
float sa = score_a, sb = score_b;
// inner loop
for (int d = 0; d < D; ++d)
{
  float smax = fmaxf(sa, sb); // repeated
  float wa   = expf(sa - smax); // repeated
  float wb   = expf(sb - smax); // repeated
  float inv  = 1.0f / (wa + wb + 1e-12f);
  float a = wa * inv, b = wb * inv;
  out[d] = a * va[d] + b * vb[d];
}
\end{lstlisting}
\caption{Baseline: recompute inside the inner loop}
    \label{fig:merge-baseline}
  \end{subfigure}
  \hfill
  \begin{subfigure}[t]{0.47\columnwidth}
\begin{lstlisting}[style=cuda-compact,basicstyle=\ttfamily\scriptsize]
// compute once per output vector
float sa = score_a, sb = score_b;
float smax = fmaxf(sa, sb);
float wa = expf(sa - smax), wb = expf(sb - smax);
float inv = 1.0f / (wa + wb + 1e-12f);
float a = wa * inv;
float b = wb * inv;

// lightweight inner loop
for (int d = 0; d < D; ++d) {
  out[d] = a * va[d] + b * vb[d];
}
\end{lstlisting}
    \caption{Optimized: hoist loop-invariant computations}
    \label{fig:merge-optimized}
  \end{subfigure}

  \caption{Hoisting loop-invariant computation in \merge.}
  \label{fig:merge-hoist}
\end{figure}

\paragraph{Kernel 2: \rmsnorm}
This kernel contains a block-level reduction that dominates runtime. As shown in \Cref{fig:reduction}, the baseline implements a tree-based reduction in on-chip shared memory, which already improves latency and bandwidth relative to a naive global-memory reduction, but progressively disables threads as the reduction proceeds. The optimized version first performs an intra-warp reduction using warp-level intrinsics (\texttt{\_\_shfl\_down\_sync}), which keeps partial sums in registers and reduces synchronization overhead. The remaining inter-warp reduction is then completed in shared memory. This register-resident intra-warp phase, followed by a short shared-memory phase, yields higher arithmetic throughput and lower memory traffic than the shared-memory-only approach.

\begin{figure}[!tb]
  \centering

  \begin{subfigure}[t]{0.47\textwidth}
    \vspace{0pt}
    \lstset{linewidth=\linewidth, xleftmargin=0pt, framexleftmargin=0pt}
\begin{lstlisting}[style=cuda-tight]
/* tx = threadIdx.x, BS = BLOCK_SIZE */
__shared__ float sm[BS];

float s = ...;        // per-thread sum
sm[tx] = s;
__syncthreads();

for (int off = BS/2; off > 0; off >>= 1) {
  if (tx < off)
    sm[tx] += sm[tx + off];
  __syncthreads();
}
...
\end{lstlisting}
    \caption{Baseline: shared-memory tree reduction}
    \label{fig:red-baseline}
  \end{subfigure}
  \hfill
  \begin{subfigure}[t]{0.47\textwidth}
    \vspace{0pt}
    \lstset{linewidth=\linewidth, xleftmargin=0pt, framexleftmargin=0pt}
\begin{lstlisting}[style=cuda-tight]
/* lane = tx & 31, warp = tx >> 5 */
float s = ...;        // per-thread sum

unsigned m = 0xffffffffu;      // intra-warp
for (int off = 16; off > 0; off >>= 1)
  s += __shfl_down_sync(m, s, off);

__shared__ float ws[BS/32];    // one per warp
if (lane == 0)
  ws[warp] = s;
__syncthreads();

...
\end{lstlisting}
    \caption{Optimized: warp-level shuffle, brief shared-memory finalize}
    \label{fig:red-optimized}
  \end{subfigure}

  \caption{Reduction strategies in \texttt{fused\_add\_rmsnorm}. \Cref{fig:red-baseline}: block-level tree reduction in shared memory with synchronization each step.
  \Cref{fig:red-optimized}: intra-warp reduction in registers using \texttt{\_\_shfl\_down\_sync}, followed by a short inter-warp aggregation in shared memory.}
  \label{fig:reduction}
\end{figure}

\paragraph{Kernel 3: \silu}

\begin{figure}[!tb]
  \centering
  \begin{subfigure}[t]{0.47\columnwidth}
\begin{lstlisting}[style=cuda-tight]
...
const __half* x_ptr = row_in;
__half xv = x_ptr[vec_idx];
...
\end{lstlisting}
\caption{Baseline: scalar half-precision}
\label{fig:base-load}
  \end{subfigure}
  \hfill
  \begin{subfigure}[t]{0.47\columnwidth}
\begin{lstlisting}[style=cuda-tight]
...
__half2* x2 = reinterpret_cast<__half2*>(row_in);
__half2 xv2 = x2[vec_idx];
...
\end{lstlisting}
\caption{Optimized: half2 vectorized load}
\label{fig:opt-load}
  \end{subfigure}
  \caption{Comparison of global-memory loads in the baseline and optimized kernels. The baseline uses a scalar half-precision load, while the optimized version employs a vectorized half2 load for improved efficiency.}
  \label{fig:vectorize}
\end{figure}

\begin{figure}[!tb]
  \centering
  \begin{subfigure}[t]{0.47\columnwidth}
\begin{lstlisting}[style=cuda-tight]
__device__ float silu_f(float x)
{
    return x / (1.0f + expf(-x));
}
\end{lstlisting}
    \caption{Baseline: standard library math + division}
    \label{fig:silu-baseline}
  \end{subfigure}
  \hfill
  \begin{subfigure}[t]{0.47\columnwidth}
\begin{lstlisting}[style=cuda-tight]
__device__ float silu_fastf(float x) {
    float y = __expf(-x);
    float r = __frcp_rn(1.0f + y);
    return __fmul_rn(x, r);
}
\end{lstlisting}
    \caption{Optimized: fast-math intrinsics}
    \label{fig:silu-optimized}
  \end{subfigure}
  \caption{Side-by-side SiLU implementations. The optimized kernel replaces a division with a reciprocal–multiply sequence and uses the fast exponential intrinsic, improving compute throughput.}
  \label{fig:fastmath}
\end{figure}

We highlight two key optimization strategies: vectorized memory access and the use of fast math intrinsics. As shown in \Cref{fig:vectorize}, the baseline kernel performs scalar loads, fetching each \texttt{\_\_half} value individually from global memory. In contrast, the optimized kernel employs vectorized loads by grouping two contiguous FP16 values into a \texttt{\_\_half2} type, allowing each instruction to retrieve a pair of elements simultaneously. This reduces the number of memory transactions and increases effective memory bandwidth. Similar vectorized access patterns are also applied in Kernel 1 and Kernel 2.

Beyond memory access, compute throughput is further improved through an optimized SiLU implementation. The baseline computes SiLU using standard math library calls and a floating-point division (\Cref{fig:silu-baseline}). The optimized kernel (\Cref{fig:silu-optimized}) instead uses CUDA device intrinsics: \texttt{\_\_expf} for exponentiation, \texttt{\_\_frcp\_rn} for reciprocal, and \texttt{\_\_fmul\_rn} for multiplication. Replacing the division with a reciprocal–multiplication sequence reduces instruction latency, improves arithmetic pipeline utilization, and achieves faster execution while preserving numerical correctness.

%% file: 6_discussion.tex
\section{Discussion}
\label{sec:discussion}

\subsection{Impact of Tensor Shapes on Performance Speedup}
\label{subsec:shape}

To study the effect of tensor shapes on performance, we report results for four representative shapes for each kernel. As shown in \Cref{tab:shape}, the kernels optimized by \name achieve consistent speedups across different shapes. For \merge (kernel 1), we use shapes of the form [seq\_len, number\_of\_heads, head\_dim]; for \rmsnorm (kernel 2) and \silu (kernel 3), we use [batch\_size, hidden\_size]. Since performance speedup varies with tensor shape, in \Cref{sec:result} we report the average speedup for each kernel across a set of common shapes drawn from widely used open-source models, ensuring that the results generalize across diverse shapes and serving scenarios.

Unlike tensor compiler optimization approaches~\cite{zheng2020ansor,zheng2022amos}, which perform shape-specific tuning, \name does not prompt agents to optimize for a particular shape. Instead, it aims to deliver performance improvements for general tensor computations.

\begin{table*}[!tb]
\centering
\small
\setlength{\tabcolsep}{6pt}
\begin{tabular}{llccc}
\toprule
\textbf{Kernel} & \textbf{Shapes} & \textbf{Time-Base ($\mu s$)} & \textbf{Time-Opt. ($\mu s$)} & \textbf{Speedup} \\
\midrule
\multirow{4}{*}{Kernel 1}
 & \texttt{[512, 32, 256]}   & 32.9 & 22.6& 1.46x \\
 & \texttt{[512, 40, 128]}   & 32.4 & 20.6 & 1.57x \\
 & \texttt{[768, 32, 256]}  & 32.5  & 32.5 & 1.00x \\
 & \texttt{[512, 64, 128]}  & 32.0 &28.2  & 1.14x \\
\midrule
\multirow{4}{*}{Kernel 2}
 & \texttt{[256, 4096]}   & 24.3 & 18.3 & 1.33x \\
 & \texttt{[1024, 4096]}   & 34.0 &  28.3 & 1.20x \\
 & \texttt{[128, 11008]}  & 25.0 & 19.4 & 1.28x \\
 & \texttt{[512, 14336]}  & 46.1 & 43.0 & 1.07x \\
\midrule
\multirow{4}{*}{Kernel 3}
 & \texttt{[16, 4096]}  & 20.9 & 14.2 & 1.47 \\
 & \texttt{[32, 5120]}  & 20.3 & 13.7 & 1.49  \\
 & \texttt{[64, 8192]} & 20.3 & 13.5 & 1.50 \\
 & \texttt{[16, 12288]} &20.4  & 13.6 & 1.50 \\
\bottomrule
\end{tabular}
\caption{Impact of tensor shapes on performance.}
\label{tab:shape}
\end{table*}

\subsection{Limitations and Future Work}
\label{subsec:limitation}

Our evaluation currently focuses on three CUDA kernels, and the framework is tailored to SGLang~\cite{zheng2025sglang}. In future work, we aim to extend support to a broader set of kernels and additional frameworks such as vLLM~\cite{kwon2023efficient}, PyTorch~\cite{paszke2019pytorch}, and TorchTitan~\cite{liang2025torchtitan}.

A key limitation is that the pre-processing and post-processing steps (\Cref{subsec:system}) are fully manual. Pre-processing requires extracting and simplifying kernels into stand-alone versions suitable as inputs to \name, while post-processing involves monkey-patching the optimized kernels back into SGLang and validating them against the original implementation. These steps are non-trivial to automate due to the complexity of modern serving frameworks. Future research should explore how to make this process more automated, potentially with human-in-the-loop guidance, so that \name{} can scale to larger sets of kernels.

%% file: 7_conclusion.tex
\section{Conclusion}
\label{sec:conclusion}

GPU kernel optimization is a critical yet labor-intensive challenge in high-performance computing and machine learning. In this work, we introduced \name, the first LLM-based multi-agent system designed specifically for GPU kernel optimization. Unlike prior approaches that translate high-level PyTorch modules into CUDA code, \name{} operates directly on existing CUDA kernels from SGLang, a widely deployed LLM serving framework. By coordinating specialized agents for code generation, testing, profiling, and planning, \name{} produces kernels that are both correct and high-performance. Our evaluation shows that \name{} delivers an average speedup of \avgspeedup, with case studies highlighting how LLMs can autonomously apply loop transformations, restructure memory access, exploit CUDA intrinsics, and leverage fast math operations. These results underscore the promise of multi-agent LLM systems as a new paradigm for kernel performance optimization.